\documentstyle[preprint,prl,aps]{revtex}
\begin{document}

\title{Acoustically driven Storage of Light in a Quantum Well}

\author{C. Rocke, S. Zimmermann, A. Wixforth, and J. P. 
Kotthaus}
\address{Sektion Physik der LMU, Geschwister-Scholl-Platz 1, D-
80539 M\"unchen, Germany}
\author{G. B\"ohm and G. Weimann\cite{Weimann}}
\address{Walter-Schottky-Institut der TUM, D-85748 Garching, 
Germany}

\date{\today}
\maketitle

\begin{abstract}
The strong piezoelectric fields accompanying a surface acoustic 
wave on a semiconductor quantum well structure are employed to 
dissociate optically generated excitons and efficiently trap the 
created electron hole pairs in the moving lateral potential 
superlattice of the sound wave. The resulting spatial separation of 
the photogenerated ambipolar charges leads to an increase of the 
radiative lifetime by orders of magnitude as compared to the 
unperturbed excitons. External and deliberate screening of the lateral 
piezoelectric fields triggers radiative recombination after very long 
storage times at a remote location on the sample. 
\end{abstract}


The dynamics of photogenerated carriers in semiconductor 
structures with reduced dimensionality has been subject of 
intensive investigations in recent years \cite{Schmitt89,Miller85}. 
State of the art band-gap engineering technologies enable us to tailor 
low-dimensional semiconductor systems with desirable 
optoelectronic properties and study fundamental aspects of carrier 
dynamics. This has increased tremendously our fundamental 
understanding of the dynamic properties of artificial semiconductor 
structures and also resulted in a wide range of novel devices such as 
quantum well lasers, modulators and detectors as well as all-optical 
switches. Nevertheless the bulk band structure of semiconductors 
seems to dominate optoelectronic properties since the strength of 
interband transitions is largely governed by the atomic-like Bloch 
parts of the wavefunction \cite{Weisbuch91}. Thus it appears at 
first glance unavoidable that strong interband optical transitions are 
linked to direct band gap semiconductors with short radiative 
lifetimes such as GaAs whereas long radiative lifetimes of 
photogenerated carriers imply utilization of semiconductors with 
indirect band gaps such as Si and correspondingly reduced interband 
absorption. Initial attempts to employ band gap engineering in 
order to combine strong interband absorption with long radiative 
lifetimes have focussed on so-called doping superlattices 
\cite{Doehler}. There alternate n- and p-doping along the growth 
direction is utilized to combine a direct gap in momentum space 
with an indirect gap in real space which causes a spatial separation 
of photogenerated  electron-hole (e-h) pairs and hence considerably 
prolonged lifetimes. 

Here, we introduce a new way of band gap engineering in which we 
expose a semiconductor quantum well of a direct gap material to a 
moving potential superlattice modulated in the plane of the well. 
We show that the confinement of photogenerated e-h pairs to two 
dimensions together with the moving lateral superlattice allows 
reversible charge separation \cite{Hoskins84}. We demonstrate that 
the combination of both the advantages of strong interband-
absorption and extremely long lifetimes of the optical excitations is 
achieved without affecting the superior optical quality of the 
quantum well material. 

The spatial separation of the electron-hole pairs is achieved via the 
piezoelectric potential of acoustic waves propagating along the 
surface of a semiconductor quantum well system. On a 
piezoelectric substrate, the elliptically polarized surface acoustic 
waves (SAWs) are accompanied by both lateral and vertical 
piezoelectric fields which propagate at the speed of sound. Those 
fields can be strong enough to field-ionize optically generated 
excitons and to confine the resulting electrons and holes in the 
moving lateral potential wells separated by one half wavelength of 
the SAW. The spatial separation dramatically reduces the 
recombination probability and increases the radiative lifetime by 
several orders of magnitude as compared to the unperturbed case. 
We further demonstrate that the dynamically trapped electron-hole 
pairs can be transported over macroscopic distances at the speed of 
sound and that deliberate screening of the lateral piezoelectric fields 
of the SAW leads to an induced radiative recombination after long 
storage times at a location remote from the one of e-h generation. 
This conversion of photons into a long lived e-h polarization which 
is efficiently reconverted into photons can serve as an optical delay 
line operating at sound velocities.

The undoped quantum well samples used in our experiments are 
grown by molecular beam epitaxy on a (100) - GaAs substrate. The 
quantum well consists of 10 nm pseudomorphic 
In$_{0.15}$Ga$_{0.85}$As grown on a 1 $\mu$m thick GaAs 
buffer and is covered by a 20 nm thick GaAs cap layer. The active 
area of the sample is etched into a 2.5 mm long and 0.3 mm wide 
mesa (see inset of Fig. 1) with two interdigital transducers (IDTs) 
at its ends. The IDTs are designed to operate at a center frequency 
$f_{SAW} = 840$ MHz. They are partially impedance matched to 
the 50 $\Omega$ radio frequency (RF) circuitry using an on-chip 
matching network thus reducing the insertion loss of each 
transducer to the 5 dB range. The sample is mounted in an optical 
cryostat and the experiments presented here are performed at 
$T=4.2$ K. Light from a pulsed laser diode ($\lambda_{laser} = 
780$ nm) is used for optical interband excitation above band gap 
and the photoluminescence (PL) of the sample is analysed in a 
triple grating spectrometer. Either a gated photomultiplier or a 
charged coupled device (CCD) serve as a detector for the PL. 
Application of a high frequency signal to one of the IDT launches a 
SAW of wavelength $\lambda_{SAW} = v_{SAW}/f_{SAW}$. It 
propagates along the [110] direction and can be detected at the 
other IDT after the acoustic delay of order of 1 $\mu$s determined 
by the spacing of the IDTs. Here, $v_{SAW} = 2865$ m/s denotes 
the SAW velocity for the given sample cut and orientation. Either 
pulsed or continous wave (cw) operation of the SAW transducers is 
possible.

In Fig. 1 we depict the 'direct', i.e. local and instantaneous PL of the 
quantum well under the influence of a SAW. Laser excitation and 
PL emission occur at the same site $x_c$ on the sample close to the 
center of the mesa with the recombination time being much shorter 
than the time resolution of our experiment (20 ns). The different PL 
traces are recorded at different acoustic power levels $P_1$ of a cw 
high frequency ($f_{SAW}=840$ MHz) signal fed into one of the 
transducers (IDT$_1$). With increasing SAW power the PL is 
shifted towards slightly higher energies and its intensity strongly 
decreases until at the highest power used it becomes completely 
quenched. This quenching is already an indirect indication of the 
increased trapping probability and subsequent transport of the e-h 
pairs, as will be demonstrated in the following.

The piezoelectric fields of the SAW modulate the band edges with 
respect to the chemical potential similar as in doping superlattices 
\cite{Doehler} or  statically imposed laterally periodic electric 
fields using an interdigitated gate electrode \cite{Schmeller94}. In 
this moving potential superlattice with period 
$\lambda_{SAW}=3.4$ $\mu$m the excitons become polarized 
predominantly by the lateral electric field until they dissociate at 
high fields into spatially separated e-h pairs. These are then 
efficiently stored in the potential minima and maxima of the 
conduction and the valence band, respectively (see inset of Fig. 1).  
For an acoustic power of $P_1=+13.5$ dBm (22.4 mW) lateral 
fields as high as $E_l=8$ kV/cm and vertical fields up to $E_v=10$ 
kV/cm are achieved \cite{Kim90,Jain89,Gryba94}. Recent related 
studies of the optical properties using statically imposed lateral 
electric fields on the same wafer \cite{Schmeller94} and our 
dynamical experiments indicate the analogy between both methods 
to create electric fields in the plane of the quantum well. The 
influence of the SAW-induced vertical fields $E_v$ via the quantum 
confined Stark effect (QCSE) can be neglected in the present case as 
we use a comparatively thin (10 nm) quantum well 
\cite{Miller85,Vina87}.

The dramatically prolonged recombination time of the trapped e-h 
pairs together with the propagation of the SAW along the surface of 
the quantum well sample enable us to directly study the transport 
of the photogenerated carriers over macroscopic distances as 
sketched in the inset of Fig. 2. Deliberate screening of the storing 
lateral potential triggers radiative recombination and thus the 
reassembly of the polarized e-h pairs into photons. In this 
experiment, optical excitation and emission of the PL do not occur 
at the same spot of the sample. Using a pulsed laser diode the 
optical excitation is accomplished at the site $x=x_{in}$, whereas 
the detector monitors the PL coming from a location $x=x_{out}$. 
The part of the sample beyond $x=x_{out}$ is covered by a thin, 
semitransparent nickel-chromium layer that readily screens the 
lateral piezoelectric fields of the SAW \cite{Wixforth89,Rocke94} 
and thus triggers recombination. At the time $t=0$ a SAW pulse of 
width $\Delta t = 200$ ns is launched at the emitting IDT$_1$. At 
around $t=t_1$ the SAW pulse is centered at $x=x_{in}$ where a 
laser pulse creates excitonic excitations. These excitons are 
immediately field-ionized and the resulting electron-hole pairs 
efficiently trapped in the moving lateral potential wells of the SAW 
as indirectly seen in Fig. 1. After an acoustic delay time of in this 
case  $\tau_{storage}=350$ ns corresponding to a transport path 
length of $x_{out}-x_{in} = 1$ mm the SAW pulse reaches the 
location $x=x_{out}$ where the lateral SAW fields are screened 
thus inducing recombination and a strong PL signal at the detector 
PM. Time delays $\tau_{storage}$ up to several $\mu$s 
corresponding to an e-h transport over some millimeters have been 
achieved without severe degradation of the transport efficiency, 
only the sample length imposing a maximum time delay.

The time-delayed PL is demonstrated in the lower part of Fig. 2. 
Here, we plot the PL intensity detected at $x=x_{out}$ as a 
function of time after the SAW excitation. The SAW and the pump 
laser pulses are indicated by the rectangles close to the ordinate. 
The slight decrease of the PL intensity at $t=t_1$ marks the arrival 
of the SAW pulse at the excitation site $x=x_{in}$ by  quenching 
the direct and instantaneous PL at $x=x_{in}$ that is detected as 
weak stray light by the photomultiplier. To avoid any possible 
spurious effect originating from this stray light we make sure that 
the pump laser is switched off well before the SAW packet arrives 
at $x=x_{out}$. About 650 ns after the launch of the SAW pulse 
we detect a strong PL at the site $x=x_{out}$ which clearly 
indicates the SAW - mediated transport and subsequent 
recombination of the photogenerated carriers. Comparison of the 
PL that is detected at the pumping site $x_{in}$ to the delayed PL 
at $x_{out}$ shows that only about 30\% of the photogenerated 
and trapped carriers are 'lost' along their way. The major 
mechanisms for this loss are believd to be non-radiative processes 
along the sound path which might be related to the high density of 
recombination centers at the mesa edges and to wave front 
inhomogeneities.

In Fig. 3, we plot the intensity of the delayed PL ($x=x_{out}$, 
$\tau > 300$ ns) as a function of the SAW power. For comparison, 
we also show the quenching of the 'direct' and instantaneous PL 
($x=x_{in}$, $\tau < 1$ ns ) as already demonstrated in Fig. 1. 
Clearly, a threshold behavior for the effective dissociation and 
transport is observed. Only for SAW powers $P_1 > -5$ dBm 
carriers are trapped by the SAW and transported to the gate 
electrode at which they recombine. At the same power level, the 
'direct' PL becomes quenched. The corresponding lateral electric 
field strength of $E_l \approx 2 \cdot 10^3$ V/cm agrees very well 
to the one expected theoretically for exciton dissociation 
\cite{Miller85}. At high power levels we observe a saturation of 
the delayed PL signal, which indicates the complete filling of the 
lateral SAW potential wells and a dynamical screening of the SAW 
fields by the photogenerated carriers.

The photon reassembly via the recombination of the stored carriers 
may also be induced following an alternative route where both the 
lifetime of the e-h pairs as well as the location of the induced 
radiative recombination may be chosen at will. In addition to the 
storing SAW$_1$ a second SAW$_2$ is launched using a second 
transducer IDT$_2$ (see Fig. 1). When colliding, the two SAWs 
interfere and create a lateral potential modulation which is 
oscillating in time and space. If both SAWs have the same 
wavelength and amplitude, the interference pattern is that of a 
standing wave. This experiment is depicted in Fig. 4. The acoustic 
power generated with IDT$_1$ is set to $P_1=+4.5$ dBm, which 
corresponds to the level at which the direct PL is already quenched 
by about 50\% (cf. Figs. 1,3). The power level $P_2$ of the 
counter-propagating SAW$_2$ of the same frequency 
$f_{SAW}=840$ MHz is varied in the experiment. As long as 
$P_1>P_2$, SAW$_1$ is effectively trapping and transporting the 
photogenerated e-h pairs. However, if $P_2$ becomes comparable 
or equal to $P_1$, both waves interfere to create a standing wave 
pattern. At this moment, the time averaged wave function overlap 
of the stored electrons and holes increases dramatically which 
results in an induced radiative recombination and a strong increase 
of the observed PL. This result clearly excludes any possible 
thermal effect that might be responsible for the quenching of the 
PL. Further increase of $P_2$ reverses the original situation as now 
SAW$_2$ takes over the charges and efficiently reduces the 
recombination probability again by a spatial separation of the e-h 
pairs. This is clearly demonstrated by the strong decrease of the PL 
intensity at high power levels $P_2>P_1$.

In summary, we demonstrate the transformation of light into an 
elementary excitation of a solid stored in the lateral superlattice 
potential of a surface acoustic wave moving at the speed of sound. 
The corresponding storage time can be orders of magnitudes longer 
than the recombination lifetime in direct bandgap semiconductors. 
Reassembly of the stored charges into photons can be induced at a 
location distant from their generation. This way, the lifetimes of the 
optical excitations in a direct semiconductor can be prolonged by 
several orders of magnitude without degradation of the superior 
optical properties of the high purity material. The possibility to 
tune the storage time of the light-generated e-h polarization by two 
counter-propagating SAWs together with the fact that the location 
of the radiative recombination on the sample can be determined 
only by a preset time delay between the two SAW pulses opens a 
wide field for novel acousto-optic devices. Optical delay, beam 
steering, multiplexing and demultiplexing of optical signals possibly 
may thus be realized on a single chip. 

We gratefully acknowledge very enlightening discussions with A.V. 
Govorov, technical advise of S. Manus, and the financial support of 
the Deutsche Forschungsgemeinschaft (DFG) and the Bayerische 
Forschungsstiftung FOROPTO.

\begin{figure}
\caption{Photoluminescence spectra of a single 10 nm wide 
InGaAs/GaAs quantum well structure for different acoustic 
powers. The optical excitation occurs at the site $x=x_c$ with an 
intensity of 10 mW/cm$^2$ and wavelenght 
$\lambda_{laser}=780$ nm. The insets schematically depict the 
sample design with two interdigital transducers and the storage of 
optically generated excitons in the potential of a surface acoustic 
wave.}
\end{figure}

\begin{figure}
\caption{Ambipolar transport of trapped charges by a SAW. At 
$t=0$ ns a 200 ns long RF pulse at $f_{SAW}=840$ MHz  
applied to IDT$_1$ generates a SAW packet with an acoustic 
power of $P_1=13.5$ dBm. At $t=t_1$ and $x=x_{in}$ the 
potential extrema of the SAW are filled with photogenerated 
electron - hole pairs which are transported with sound velocity to a 
semitransparent metallization at $x=x_{out}$. Here the deliberate 
screening of the piezoelectric potential modulation lifts the spatial 
separation of the carriers and induces radiative recombination at 
$x=x_{out}$ and $t=t_2$. The duration of the RF pulse and the 
laser pulse are indicated in the lower part.}
\end{figure}

\begin{figure}
\caption{Normalized direct and delayed PL intensities as a function 
of the acoustic power $P_1$. Efficient ambipolar transport of 
charges resulting in a time-delayed PL signal requires a minimum 
threshold acoustic power of $P_1=-5$ dBm which corresponds to a 
minimum lateral electric field strengh ($E_l=2$ kV/cm). This 
threshold coincides with the quenching of the direct PL signal.}
\end{figure}

\begin{figure}
\caption{Direct PL intensity at $x=x_c$ as a function of the 
acoustic power $P_2$ for a constant acoustic power $P_1=4.5$ 
dBm. For the standing wave geometry at $P_1=P_2$ the spatial 
separation between electrons and holes is lifted resulting in a 
recovered PL intensity due to the increased transition probability.}
\end{figure}

\end{document}